\documentclass[12pt,preprint]{aastex}
\usepackage[]{natbib}
\usepackage[]{amsmath}
\usepackage{graphicx}  

\newcommand{\myemail}{klement,rix@mpia.de}
\newcommand{\fuchsemail}{fuchs@ari.uni-heidelberg.de}

\shorttitle{Identifying Stellar Streams in the 1$^{st}$ RAVE Public Data Release}
\shortauthors{Klement et al.}

\begin{document}

\title{Identifying Stellar Streams in the 1$^{st}$ RAVE Public Data Release}

\author{R. Klement\altaffilmark{1}, B. Fuchs\altaffilmark{2} and H. W. Rix\altaffilmark{1}}

\altaffiltext{1}{Max-Planck-Institut f\"ur Astronomie, K\"onigstuhl 17, D-69117 Heidelberg; \myemail}
\altaffiltext{2}{Astronomisches Rechen-Institut am Zentrum f\"ur Astronomie Heidelberg, M\"onchhofstraße 12-14, D-69120 Heidelberg; \fuchsemail}

\begin{abstract}
We searched for and detected stellar streams or moving groups in the solar neighbourhood, using the
data provided by the 1$^{st}$ RAVE public data release. This analysis is based on distances to RAVE stars
estimated from a color-magnitude relation that was calibrated on Hipparcos stars. Our final sample consists of 7015 stars selected to be within 500 pc of the Sun and to have distance errors better than 25\%. Together with radial velocities from RAVE and proper motions from various data bases, there are estimates for all 6 phase-space coordinates of the stars in the
sample. We characterize the orbits of these stars through suitable proxies for their angular momentum and eccentricity, and compare the observed distribution to the expectations from a smooth distribution. On this basis we identify at least four "phase space overdensities" of stars on very similar orbits in the Solar neighbourhood. We estimate the statistical significance of these overdensities by Monte Carlo simulations.
Three of them have been identified previously: the Sirius and Hercules moving group and a stream found independently in 2006 by Arifyanto and Fuchs and Helmi et al.
In addition, we have found a new stream candidate on a quite radial orbit, suggesting an origin external to the Milky Way's disk. Also, there is evidence for the Arcturus stream and the Hyades-Pleiades moving group in the sample. The detections are further supported by analysing the stellar distribution in velocity and angular momentum space using the same Monte Carlo simulations. We also find that the significance of overdensities is comparable, independent of the space in which the stream search is conducted. This analysis, using only a minute fraction of the final RAVE data set, shows the power of this experiment to probe the phase-space substructure of stars around the Sun. 
\end{abstract}

\keywords{Galaxy: solar neighbourhood --- Galaxy: kinematics and dynamics}

\defcitealias{ari06}{AF06}
\section{Introduction}

According to current ideas about the cosmogony of galaxies these were assembled
through hierarchical merging, which should result in a richly structured phase-space distribution of dark matter and stars. Direct emperical evidence
for such events has been sought among the stellar populations of the Milky Way with large scale surveys. Ongoing satellite accretion events have
been discovered in several instances. A prominent
example is the Sagittarius galaxy \citep{iba94} with its associated
tidal stream which wraps around the Galaxy nearly perpendicular to the
galactic plane (\citet{iba01}, \citet{maj03}, \citet{bel06}). Simmilarly the Monoceros
stream at low galactic latitudes \citep{yan03,iba03} or
the recently discovered Orphan stream \citep{bel06} are interpreted
as tidal debris from the Canis Major and Ursa Major II dwarf galaxies,
respectively \citep{pen05,fell07}. Numerical simulations have shown that such debris streams can survive as coherent structures over gigayears \citep{helm03,pen05,law05}.

Star streams, i.e. goups of stars on essentially the same orbits in the
galactic potential, have also been detected as overdensities in the phase
space distribution of stars in the Solar neighbourhood. The concept of
"moving groups" originates from the work of Eggen
\citep[ and references therein]{egg96}. Some of the moving groups are
associated with young open clusters and can be naturally explained as
clouds of former members, now unbound and drifting away from the clusters; these moving groups only reflect the nature of star formation, not necessarily that of hierarchical galaxy formation.

However, data from the last decade seem to support the concept of cold star streams in the Solar neighbourhood, consisting of old stars (5-10 Gyrs).
\citet{helm99} discovered the signature of a cold stream in the velocity
distribution of halo stars which had been constructed from
Hipparcos data. This was later confirmed by \citet{chi00} using their
own data. \citet{helm99} argued that this stream formed part of the
tidal debris of a disrupted satellite galaxy accreted by the Milky Way, which
ended up in the halo. Moreover, \citet{nav04} interpreted Eggen's
Arcturus stream as such another debris stream, but in the thick disk of the
Milky Way, dating back to an accretion event 5 to 8 Gyrs ago.

Moving groups of old stars are also observed in the velocity distribution of thin
disk stars in the Solar vicinity. Based on Hipparcos parallaxes and
proper motions, \citet{dehn98} found with statistical methods new evidence
for the Sirius-UMa, Pleiades-Hyades, and Hercules star streams. Even more
convincingly these streams show up in the three-dimensional kinematical data
of F and G stars in the Copenhagen-Geneva survey \citep{nord04}.
These moving groups are very probably not related to tidal debris streams,
but originated from dynamical effects within the disk itself like
resonances with the inner bar of the Milky Way \citep{deh00} or with spiral
denisty waves \citep{qui05}.

Yet another stream has been discovered independently by \citet[hereafter AF06]{ari06} and \citet{helm06}
analyzing different data samples and applying different detection techniques.
The metallicities of the stars in this stream cover a broad range,
from metallicities typical for old thin disk stars to that of thick disk stars.
Thus the exact nature of this stream seems to be at present not fully
understood.

With the first data release from the RAVE collaboration \citep{stei06}
a new large data sample of stars in the Milky Way became available, which is
ideally suited for kinematical studies. To explore RAVE's potential, we analyze
the velocity distribution of more than 7000 stars within a distance of 500 pc
and search for overdensities in phase space using the projection technique of
\citetalias{ari06}. Even the first data release sample is so substantive that we can determine
signal-to-noise ratios for orbital 'overdensities' and show the statistical significance of the
detected overdensities.

\section{Data}
\label{data}

The RAVE DR1 \citep{stei06} contains 25,274 radial velocities for 24,748 individual stars together with proper motions and photometry from other major catalogs (Starnet 2.0, Tycho-2, SuperCOSMOS, USNO-B, DENIS and 2MASS). Its total sky coverage is $\sim4760$ deg$^2$. The only missing parameter for getting all velocity and position components is a distance estimate. The photometric data, however, allow the application of a photometric parallax relation\footnote{We assume that the vast majority of RAVE stars are main sequence stars, because giants in the RAVE magnitude range would lie so far away, that the star density is substantially decreased. See also the discussion in \citet{sieg02}}. To calibrate such a relation for RAVE stars, we used main sequence (MS) stars (luminosity class V) from the Hipparcos catalog \citep{esa97} with accurately known parallaxes ($\frac{\sigma_\pi}{\pi}<0.1$) that could also be identified in the Tycho-2, USNO-B and 2MASS data. These other catalogs contain B1, V$_T$ and H-band magnitudes\footnote{We do not consider very metal-poor (sub)dwarfs, which can lie up to $\gtrsim1$ magnitude below the Solar-metallicity MS; for metal-poor stars we would overestimate the distance, and hence velocity. The effect for disk stars, however, will be negligible, because they exhibit approximately Solar metallicity. We will come back to the subdwarf problem later in the context of the detection of a new halo stream in the RAVE data.}.

The absolute magnitudes of the stars were calculated through
\begin{equation}\label{distmod}
	M=m-10+5\log\frac{\pi}{mas}
\end{equation}

\begin{figure}[htb]
\epsscale{1.0}
\plotone{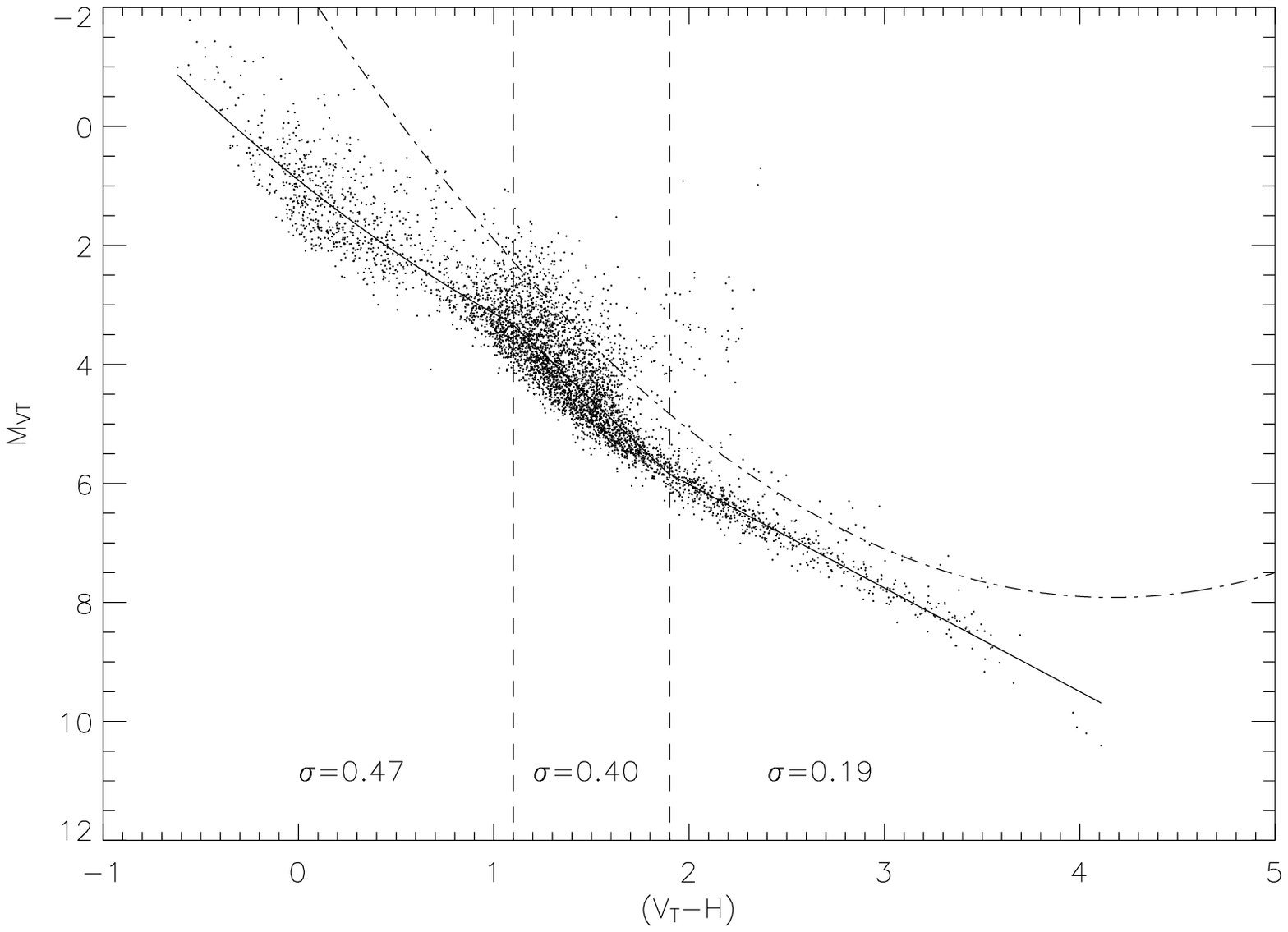}
\caption{Absolute magnitude calibration for Hipparcos MS stars with a parallax accuracy better than 10\% with $V_T-H$ as the color. All stars above the dash-dotted line are assumed to be giants incorrectly classified as MS stars and are not considered for the color-magnitude fit. The solid line shows the $M_{V_T}-(V_T-H)$ relation adopted, and the dispersions $\sigma$ about this mean relation are indicated in three color regimes.}\label{VTminH}
\end{figure}
We then considered two color-magnitude relations: $(V_T-H)$ vs. $M_{V_T}$ and $(B1-H)$ vs. $M_{B1}$. Figure~\ref{VTminH} shows the relation $M_{V_T}$ vs. $(V_T-H)$. The dash-dotted line was chosen to remove all stars which seemed to be mis-classified as MS stars in the Hipparcos catalog. This cut was done by eye, since one could only reliably distinguish a (sub-)giant from a MS star in the vicinity of the MS by measuring its surface gravity. Next, the color-magnitude diagram was divided into three color bins in which we seperately fitted a color-magnitude relation (solid lines). For each bin, the intrinsic scatter of the color-magnitude relation was calculated as the 1/2 of the central 68\% of the cumulative distribution of the differences $\Delta M$ between the true absolute magnitudes (obtained through the parallax) and those obtained through the fit. The scatter in each bin is denoted in Figure~\ref{VTminH}. 
\begin{figure}[htb]
\epsscale{1.0}
\plotone{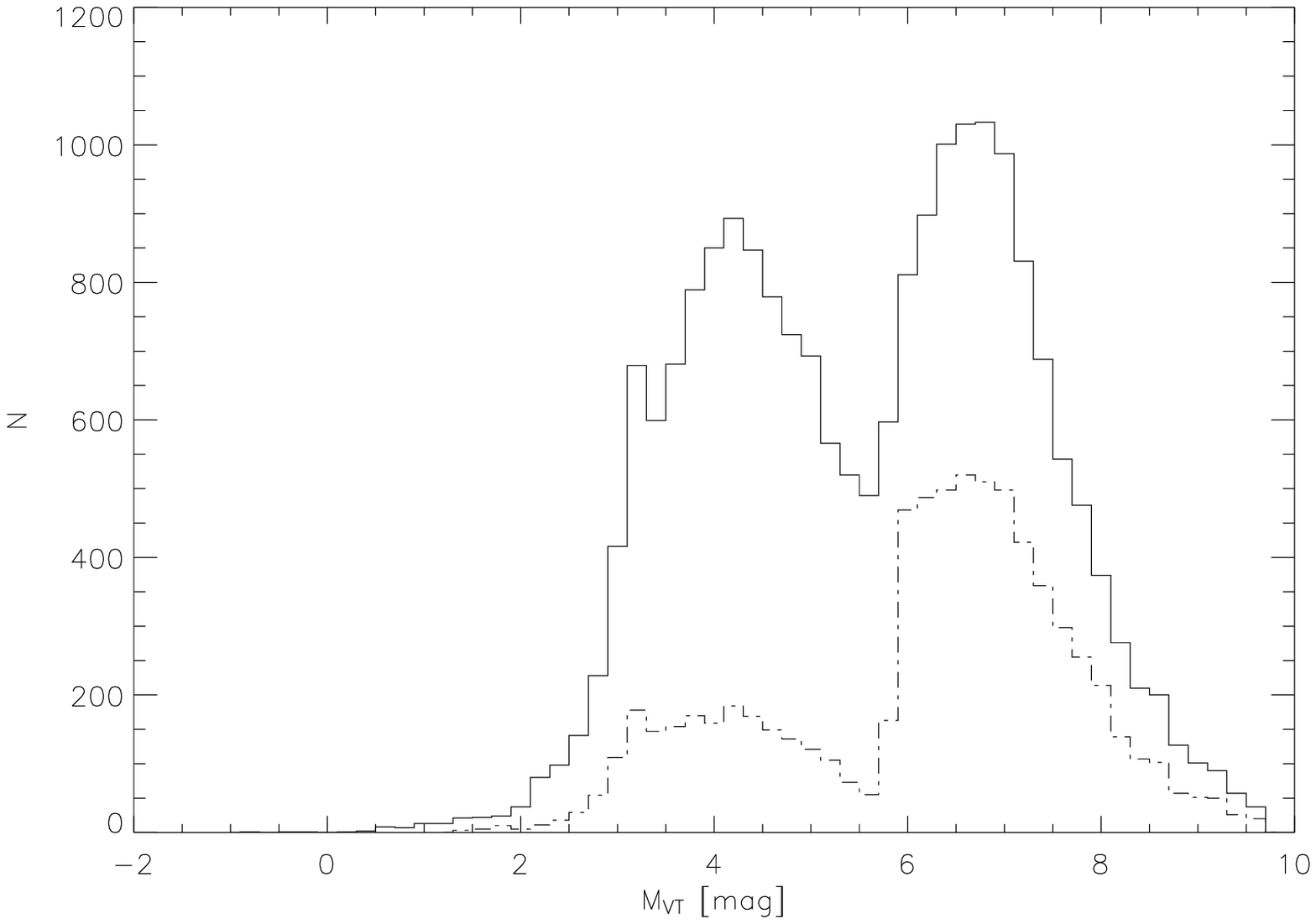}
\caption{Distribution of absolute magnitudes of all DR1 RAVE stars (solid line) and our selected sample (dash-dotted line). A large fraction has $M_{V_T}>6$.}\label{Mvthist}
\end{figure}
For our analysis we adopted the $M_{V_T}$ vs. $(V_T-H)$-relation, because its intrinsic scatter is slightly smaller than that of $M_{B1}$ vs. $(B1-H)$. Also, the errors of $B1$ for each star in the RAVE catalog are unknown, while the errors in $V_T$ are given. The formal error in $H$ (from 2MASS) is typically very small, $\sim0.01$ mag. For that reasons we chose the $M_{V_T}(V_T-H)$-relation to calculate absolute magnitudes $M_{V_T}$ for the RAVE stars. If $V_T$ was not given for a RAVE star, it was possible for most of the stars to get this magnitude through conversion formulas from other magnitudes like the USNO-B $B1$ and $R1$ (Klement, unpublished). Figure~\ref{Mvthist} shows that a large fraction of all DR1 stars has absolute magnitudes fainter than $M_{V_T}=6$, i.e. lie in the section of the color-magnitude relation where its intrinsic scatter is smallest. Therefore we expect to get a large sample of stars with distance estimates good to 10\% - 20\% in most cases (see below and Figure~\ref{Mvthist}, where the $M_{V_T}$-distribution of our final sample is shown).

We established a right-handed Cartesian coordinate system $(x,y,z)$ centered on the Local Standard of Rest (LSR) with the $x$-axis pointing in the direction of the Galactic center, the $y$-axis pointing in the direction of Galactic rotation and the $z$-axis in the direction of the North Galactic Pole. $(U,V,W)$ denote the velocity components of a star in this Cartesian coordinate system. The Sun is assumed to lie at a distance $R_\odot=8kpc$ from the Galactic center and to move with velocity $(U_\odot,V_\odot,W_\odot)$=(10.0,5.2,7.2)km s$^{-1}$ with respect to the LSR \citep{deh98}. A star's distance $d$ follows from the apparent and absolute magnitudes $(V_T,M_{V_T})$ via Equation \ref{distmod}, with its error $\sigma_d$ determined through:
\begin{equation}
	\sigma_d=\frac{1}{5}d\ln10\sqrt{(\sigma_{V_T})^2+(\sigma_{M_{V_T}})^2},
\end{equation}
where $\sigma_{V_T}$ and $\sigma_{M_{V_T}}$ denote the errors of $V_T$ and $M_{V_T}$ respectively. The main error is introduced by the intrinsic scatter of the color-magnitude relations (Figure~\ref{VTminH}), which is translated into $\sigma_{M_{V_T}}$.

The velocity components of a star can now be calculated from its position, radial velocity $v_{rad}$ and proper motions $(\mu_\alpha,\mu_\delta)$ through formulae given e.g. in Section III of \citet{joh87}:
\begin{eqnarray}
	\Biggl(
\begin{array}{c}
  U\\
	V\\
	W
\end{array} \Biggr)=\Biggl(
\begin{array}{c}
  U_\odot\\
	V_\odot\\
	W_\odot
\end{array} \Biggr)+\textbf{B}
\Biggl(
\begin{array}{c}
  v_{rad}\\
	4.74\:\mu_\alpha\: d\\
	4.74\:\mu_\delta\: d
\end{array} \Biggr)
\end{eqnarray}

\begin{figure}[htb]
\epsscale{1.0}
\plotone{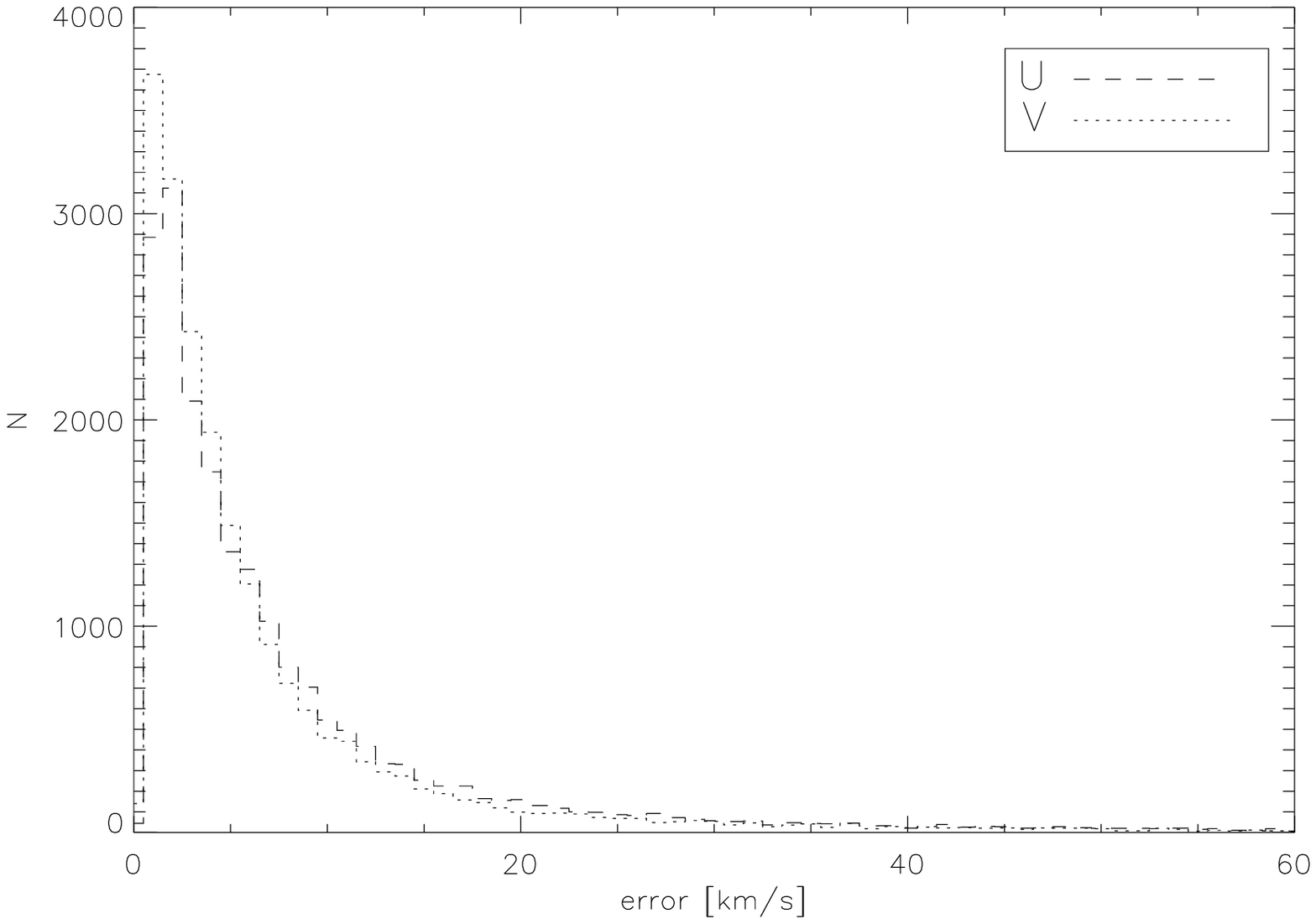}
\caption{Distribution of errors for the velocity components U (dashed) and V(dotted) for all stars in RAVE DR1 as derived from Equation~\ref{error-estimates}. These errors include radial velocity errors, proper motion errors and distance uncertainties. Both distributions have a peak around 2-3 km s$^{-1}$ and a long tail reaching to much higher values.}\label{vel-err-dist}
\end{figure}

If $\mu_\alpha$ and $\mu_\delta$ are taken in units of [mas yr$^{-1}$] and $d$ in [kpc], the factor 4.74 gives the result in [km s$^{-1}$]. The coordinate transformation matrix \textbf{B} is defined in \citet{joh87} and we evaluated it for the epoch 2000. The uncertainties in the velocity components depend on the uncertainties in radial velocity, proper motions and distance \citep{joh87}:
\begin{eqnarray}\label{error-estimates}
	\Biggl(
\begin{array}{c}
  \sigma_U^2\\
	\sigma_V^2\\
	\sigma_W^2
\end{array} \Biggr)=\textbf{C}
\Biggl(\,
\begin{array}{c}
  \sigma_{v_{rad}}^2\\
	(4.74d)^2\bigl[\sigma_{\mu_\alpha}^2+(\mu_\alpha\sigma_d/d)^2\bigr]\\
	(4.74d)^2\bigl[\sigma_{\mu_\delta}^2+(\mu_\delta\:\sigma_d/d)^2\bigr]
\end{array} \Biggr) \nonumber\\ 
+\:2\mu_\alpha\mu_\delta(4.74\sigma_d)^2\Biggl(
\begin{array}{c}
b_{12}\cdot b_{13}\\
b_{22}\cdot b_{23}\\
b_{32}\cdot b_{33}
\end{array}\Biggr)
\end{eqnarray} 
The elements of the matrix \textbf{C} are the squares of the corresponding elements of the matrix \textbf{B}, i.e. $c_{ij}=b_{ij}^2$ for all $(i,j)$. About 80\% of the stars in DR1 have radial velocity accuracies better than 3.4 km s$^{-1}$ and 69.1\% have a mean proper motion error of at most 2.6 mas yr$^{-1}$ \citep{stei06}. If we restrict our sample to stars within $d_{max}=500 pc$ from the Sun, i.e. within a volume of $V_{max}=\frac{4\pi}{3}d_{max}$, we can assume a typical distance for a star as the distance where the volume is half as large: $<d>=0.5^{1/3}d_{max}\approx400$ pc. If we further restrict the sample to stars with relative distance errors of $\frac{\sigma_d}{d}\leq0.25$ and assume a typical proper motion of 15 mas yr$^{-1}$, we get as an estimate for the uncertainty of the transverse velocity $\lesssim$8.6 km s$^{-1}$. The velocity error distribution in U and V is shown in Figure~\ref{vel-err-dist}. It is peaked at an error smaller than 5 km s$^{-1}$.

For the kinematic stream search we selected only those stars which satisfied the following criteria: $\frac{\sigma_d}{d}\leq0.25$, $d\leq500pc$, a total space velocity of $|\vec{v}_{tot}|\leq$ 350 km s$^{-1}$, and $\sigma_U,\sigma_V \leq 35$ km s$^{-1}$. We also restricted ourselves to stars with $l>200°$, $b>20$, because in this field the observation density was highest. This leaves 7015 stars, among which we searched for signatures of local stellar streams.

\section{Search strategy for streams}\label{SSS}

The distribution of stellar streams in velocity space is not trivial and depends on the origin and the age, or type, of the stream stars \citep{dehn98,fam04}. Principally, one has to distinguish between two types of streams: dynamical streams, i.e. groups of stars that are trapped in a small region of phase space by dynamical resonances, and streams, which stars originated from the same bound object, like a cluster or a satellite galaxy. Once those stars become unbound, they will have slightly different orbits, and hence frequencies, so that they will phase-mix. This leads to a broadening in the velocity distribution, which is more prominent in the $W$-component, because the vertical frequency is shorter than the horizontal frequencies. These streams will show a typical "banana" shaped distribution in $U$ and $V$ and a symmetric distribution in $U$ and $W$ \citep{helm06}. The distribution in $V$ of nearby stream stars with similar angular momenta is narrow, because these stars have basically the same azimuth. On the other hand, dynamical streams show a more clumplike structure in $U$ and $V$. Their location is primarily set by the tangential velocity component $V$, because $V$ is a measure for the guiding radius of a star in the Solar neighbourhood, which itself is a measure for the location of Lindblad resonances with a periodic perturbation \citep{bt87,qui05}. They also show a broad range of $W$ velocities due to phase-mixing \citep{dehn98}. Generally, scattering processes and measurement errors tend to symmetrize and broaden the velocity distribution of stellar streams of both types and makes it difficult to identify them in velocity space, particularly if the stream is very widespread spatially and consists only of a small fraction of stars compared to the smooth background.

According to \citet{helm99} and \citet{helm00} a better approach to search for streams is in the space spanned up by the integrals of motion - i.e. the energy $E$ and angular momenta $L_z$, $L_\perp=\sqrt{L_x^2+L_y^2}$ or $L_{total}$. For example, \citet{helm99} detected two fossil streams originating from the same progenitor as a clump of solar neighbourhood halo stars in $(L_z,L_\perp)$ space. Using cosmological N-body simulations, \citet{choi07} showed that over 8 Gyr the overall position of a disrupting satellite and its tidal tails basically remains the same in ($E,L_z$) space, but the shape of the distribution shifts and one satellite can produce several apparently disassociated clumps. This, together with the fact that the total angular momentum is not really an integral of motion, can obscure the signature of a well defined moving group in phase space. Therefore, \citet{helm06} proposed to look for stellar streams in a space spanned up by apocenter, pericenter and angular momentum $L_z$. Moving groups cluster around lines of constant eccentricity. 
 
We follow a similar approach in the sense that we try to find stars with the same orbital eccentricity. Our strategy for finding nearby stellar streams in velocity space is based on the Keplerian approximation for orbits developed by \citet{dek76} and summarized in \citetalias{ari06}. We assume an axisymmetric potential and that stars in the same stellar stream move on orbits that stay close together, which is justified by numerical simulations of satellite disruption \citep{helm06}. These stars should form a clump in the projection of velocity space spanned up by $\sqrt{U^2+2V^2}$ and $V$. The latter is related to the angular momentum $L_z$ which defines the guiding center orbits of the stars. The first quantity is a measure for a star's eccentricity $e$. For a flat rotation curve $e$ is given by
 \begin{equation}
	e=\sqrt{\frac{U^2+2V^2}{2V_{LSR}^2}}.
\end{equation}
$V_{LSR}$ denotes the circular velocity of the Local Standard of Rest for which we take the IAU standard value 220 km s$^{-1}$.

Also, it can be shown that the radial action integral $J_R$ in the Keplerian approximation is approximately equal to $\frac{\pi R_\odot}{2V_{LSR}}(U^2+2V^2)$ (Fuchs, private communication). Therefore, the quantity $\sqrt{U^2+2V^2}$ should be robust against slow changes in the potential.

For the current investigation we do not include W velocity components in our search for local streams, because - as mentioned above - strong phase-mixing will smear out any coherent feature over short timescales. However, in the appendix Section~\ref{App2} we will make use of the $W$ velocities when we compare our method to the more traditional searches in $(U,V,W)$ and $(L_z,L_\perp)$ space.

\section{Results and Discussion}\label{RaD}
We now use the distribution of the 7015 selected RAVE stars in $\sqrt{U^2+2V^2}$ vs. $V$, shown in Figure~\ref{scatterplot}, to search for substructure in the kinematic distribution. Figure~\ref{scatterplot} also shows the error ellipses for some points; most of the relative errors are rather small and even though some of the larger errors produce an uncertainty in the exact position of the data points and tend to smear out substructure, several suggestive "overdensities" of stars are visible by eye. In this section we focus on identifying overdensities and on quantifiying their significance.

\begin{figure}[htb]
\caption{Distribution of our sample of RAVE stars in $\sqrt{U^2+2V^2}$ vs. $V$. We also show error ellipses for a small subset of stars; note that while the large error ellipses are most prominent, most ellipses are very small.}\label{scatterplot}
\end{figure}

\subsection{The Wavelet Transform}
To this end, we follow the same procedure outlined in \citetalias{ari06} and use the wavelet-transform technique using a two-dimensional analysing wavelet $\Psi(x,y)$. We bin the data in pixels of 2 km s$^{-1}$ width on each side and calculate the value of the wavelet-transform in each bin through:
\begin{eqnarray}\label{wl}
	w(x,y)&=&\int\int dx'dy'\Psi(x-x',y-y')\times\nonumber\\
	& &\sum_{i=0}^{N-1}\delta(x'-x_i)\delta(y'-y_i)\nonumber\\
	&=&\sum_{i=0}^{N-1}\Psi(x-x_i,y-y_i),
	\end{eqnarray}
where $N=7015$ is the number of stars in our sample.
	
Motivated by the work of \citet{sku99}, \citetalias{ari06} used a Mexican hat shaped kernel function to detect overdensities in $\sqrt{U^2+2V^2}$ vs. $V$. However, since the former quantity is not uncorrelated to the latter, we expect clumps that would be roughly spherical in, e.g., $U$ vs. $V$, to be elongated along $U=0$. That means that the performance of a Mexican hat shaped kernel is sub-optimal and might influence the significance of the overdensities. Therefore, we build a kernel function that is elongated about a factor $q$ along the V-shaped $U=0$ lines in the following way:

We rotate the coordinate axes such that the new $V$-axis, $V'$, lies along the line $U=0$ and the new $\sqrt{U^2+2V^2}$ axis perpendicular to $V'$. For simplicity we rename $V$ as $x$ and $\sqrt{U^2+2V^2}$ as $y$. This implies two different rotations, depending on wether the features are elongated in the region $V\leq0$ (clockwise rotation by the angle $\varphi=\arccos \sqrt\frac{1}{3}$) or in the region $V>0$ (counter-clockwise rotation by $\varphi$). Based on the Mexican hat shaped function, we then express the analysing wavelet as a function of the new coordinates $(x',y')$ and elongate it along the $x'$-axis by a factor $q$:
\begin{equation}\label{anawavelet}
\Psi(x',y')=\Bigl(2-\frac{x'^2}{(qa)^2}-\frac{y'^2}{a^2}\Bigr)\exp\Bigl(-\frac{x'^2}{2(qa)^2}-\frac{y'^2}{2a^2}\Bigr).
\end{equation}
This function is normalized in the sense that it's volume integral is zero. The scale parameter $a$ is a measure for the extend of the "bumps".
With the transformation equations 
\begin{subequations}
\begin{align}
x'&=\sqrt{\frac{1}{3}}x\mp\sqrt{\frac{2}{3}}y\\
y'&=\pm\sqrt{\frac{2}{3}}x+\sqrt{\frac{1}{3}}y,
\end{align}
\end{subequations}
where the upper sign stands for the case $V\leq 0$, 
it's straightforward to calculate the kernel in the unrotated coordinate system:
\begin{eqnarray}
\Psi(x,y)&=&\Bigl(2-\frac{1}{3(qa)^2}\bigl((1+2q^2)x^2+(2+q^2)y^2\mp(1-q^2)2\sqrt{2}xy\bigr)\Bigr)\times\nonumber\\& &\exp\Bigl(-\frac{1}{6(qa)^2}\bigl((1+2q^2)x^2+(2+q^2)y^2\mp(1-q^2)2\sqrt{2}xy\bigr)\Bigr)
\end{eqnarray}
Then Equation~\eqref{wl} can be used to calculate the value of the wavelet transform in each bin. 

\begin{figure}[htb]
\epsscale{1.0}
\plotone{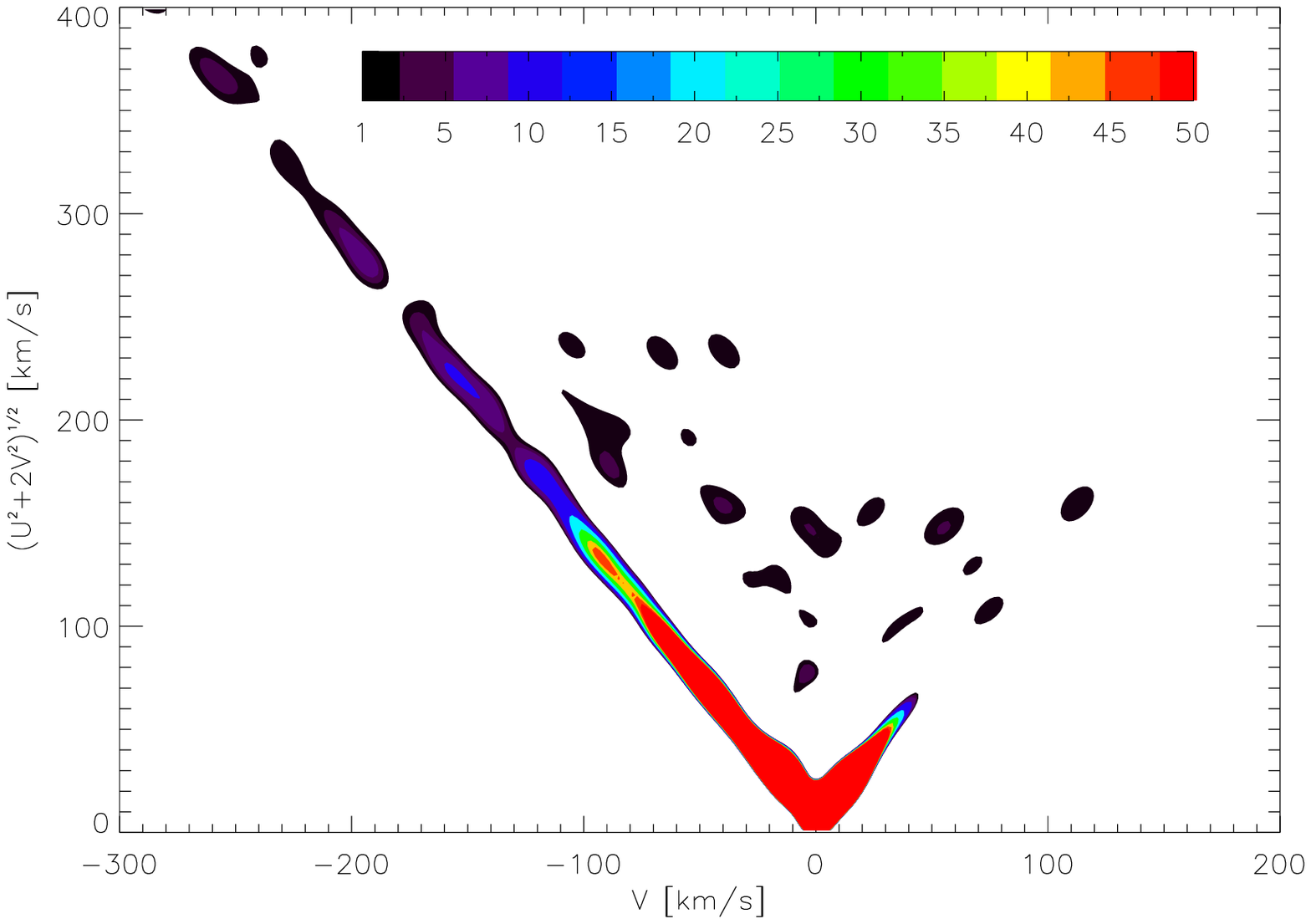}
\caption{Contours of the wavelet transform of the distribution of our sample of RAVE stars (Equation \ref{wl}) in $\sqrt{U^2+2V^2}$ vs. $V$. The contour levels are displayed in the color bar, the scale parameter of the analysing wavelet is 10 km s$^{-1}$.}\label{ker}
\end{figure}

For the elongation parameter $q$ we used $\sqrt{3}$, because the projection of a certain range in $V$ along the line $U=0$ is $\sqrt{3}$ times as long. For the scale parameter $a$ we chose 7 km s$^{-1}$, comparable to the mean errors, which extracts the overdensities most clearly (Figure~\ref{ker})\footnote{We also tested different choices for the set $(q,a)$, like $(\sqrt{2},8$km s$^{-1})$ and basically got the same results.}.

Figure~\ref{ker} shows that the bulk of our sample stars have thin-disk-like kinematics, with $|U|$ and $|V|\lesssim20$ km s$^{-1}$. The contour levels do not reflect some finer structures at $-80\lesssim V\lesssim20$ km s$^{-1}$. In this range we would expect the Hercules ($V\approx -50$ km s$^{-1}$) as well as the Hyades-Pleiades ($V\approx-20$ km s$^{-1}$) and Ursa Major/Sirius ($V\approx +4$ km s$^{-1}$) streams \citep{fam04}. Indeed, the distribution in $(\sqrt{U^2+2V^2},V)$ is somewhat bulged at these velocities.

Among the stars with high orbital eccentricities, presumably thick disk or halo stars, Figure~\ref{ker} appears to show five different overdensities. The first one at $V\approx -95$ km s$^{-1}$ is most probably the stream which was discovered independently by \citetalias{ari06} and \citet{helm06}. The second overdensity at $V\approx -120$ km s$^{-1}$ has the same kinematics as the Arcturus group discovered by \citet{nav04}. These two streams have most probably an origin external to the Milky Way disk, because their phase-space distribution resembles that of simulated accreted dwarf sattelites \citep{helm06}. However, nothing is known about their progenitors yet. The third feature at $V\approx -160$ km s$^{-1}$ has to our knowledge not yet been described in the literature. The same is true for the two clumps at $V\approx -200$ km s$^{-1}$ and $V\approx +50$ km s$^{-1}$. However, at these velocities our sample is very sparsely populated, so that just a few stars are enough to create a high value of the wavelet transform. Every isolated star in the middle of a bin increases the value of the wavelet in this bin by 2. This also explains the other clumps lying off the V-shaped main feature in Figure~\ref{ker}.

\subsection{Subtracting a Smooth Velocity Distribution}\label{Analysis}
\begin{figure}[!htb]
\epsscale{1.0}
\plotone{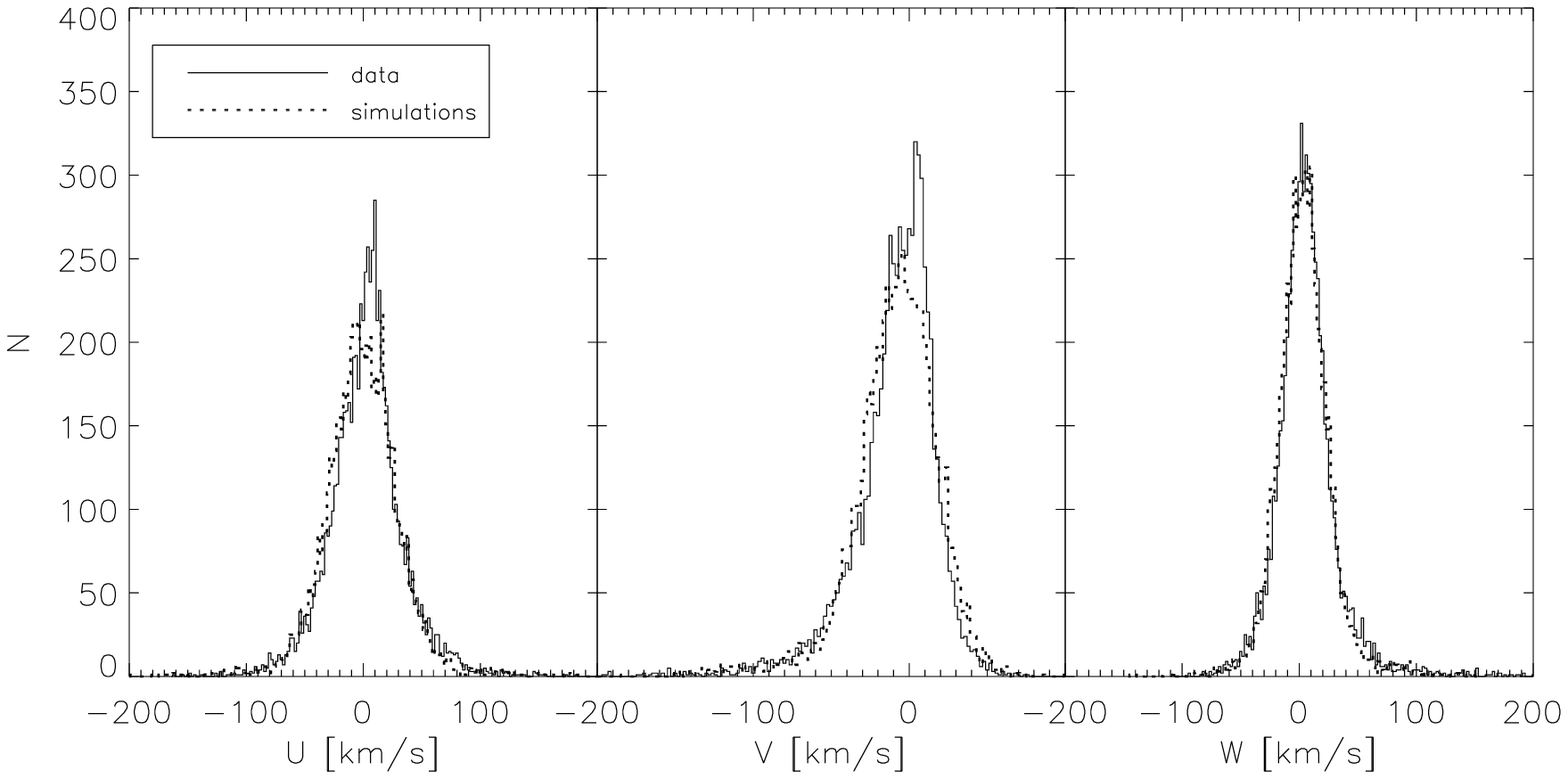}
\caption{Velocity distributions of our selected RAVE stars compared to one MC realization. The MC sample consists of three Schwarzschild distributions for the thin and thick disk and the halo, respectively. Note the peak in the data at around $V=+4$ km s$^{-1}$, which we later interpret as the Sirius moving group.}\label{vel-hist}
\end{figure}

To test whether any of these kinematic overdensities, or streams, reflected as peaks or clumps in the wavelet transform, are significant, we performed 250 Monte Carlo (MC) simulations of the same number of stars as in our sample, which we randomly draw from a Galactic model consisting of three Schwarzschild distributions \citep{bt87} to represent the thin and thick disk and the halo. The goal was to create a "smooth" reference model velocity distribution that matches the overall velocities of the RAVE sample. To this distribution, we then added normally disributed velocity errors, based on the observed error distributions shown in Figure~\ref{vel-err-dist}. We chose a local thick-to-thin disk and halo-to-thin disk normalization of 0.1 and 0.001, respectively, in agreement with the value from \citet{chen01,jur08} and the upper limit given by \citet{sieg02}. The thin and thick disk and halo are assumed to have velocity dispersions ($\sigma_U,\sigma_V,\sigma_W$) and rotational offsets from the LSR equal to (25,21,17,$-5$), (74,50,50,$-44$) and (189,97,100,$-219$) km s$^{-1}$, respectively. The values for the thin disk have been chosen to best match the smooth part of the observed velocity distribution (Figure~\ref{vel-hist}), although the ratio $\sigma_U/\sigma_V$ in the thin disk should be approximately 1.6 \citep{deh98}. Also, we chose an offset in the $W$-distribution of +3 and +10 km s$^{-1}$ for the thin and thick disk to get a good match\footnote{This deviation from the expected mean value of $\overline{W}=0$ can be due to the limited sky coverage of our sample}. The matches shown in Figure~\ref{vel-hist} are not perfect, but the slope of the velocity distribution at negative V velocities is reproduced well in the simulation. The distributions differ most clearly around $V=0$ km s$^{-1}$, because - possibly due to the Sirius moving group - the RAVE stars are peaked around $V\approx$ +5 km s$^{-1}$. Moreover, the assumption of a Schwarzschild distribution does not reflect the skewness of the velocity distribution due to the asymmetric drift effect.

\begin{figure}[!htb]
\epsscale{1.0}
\plotone{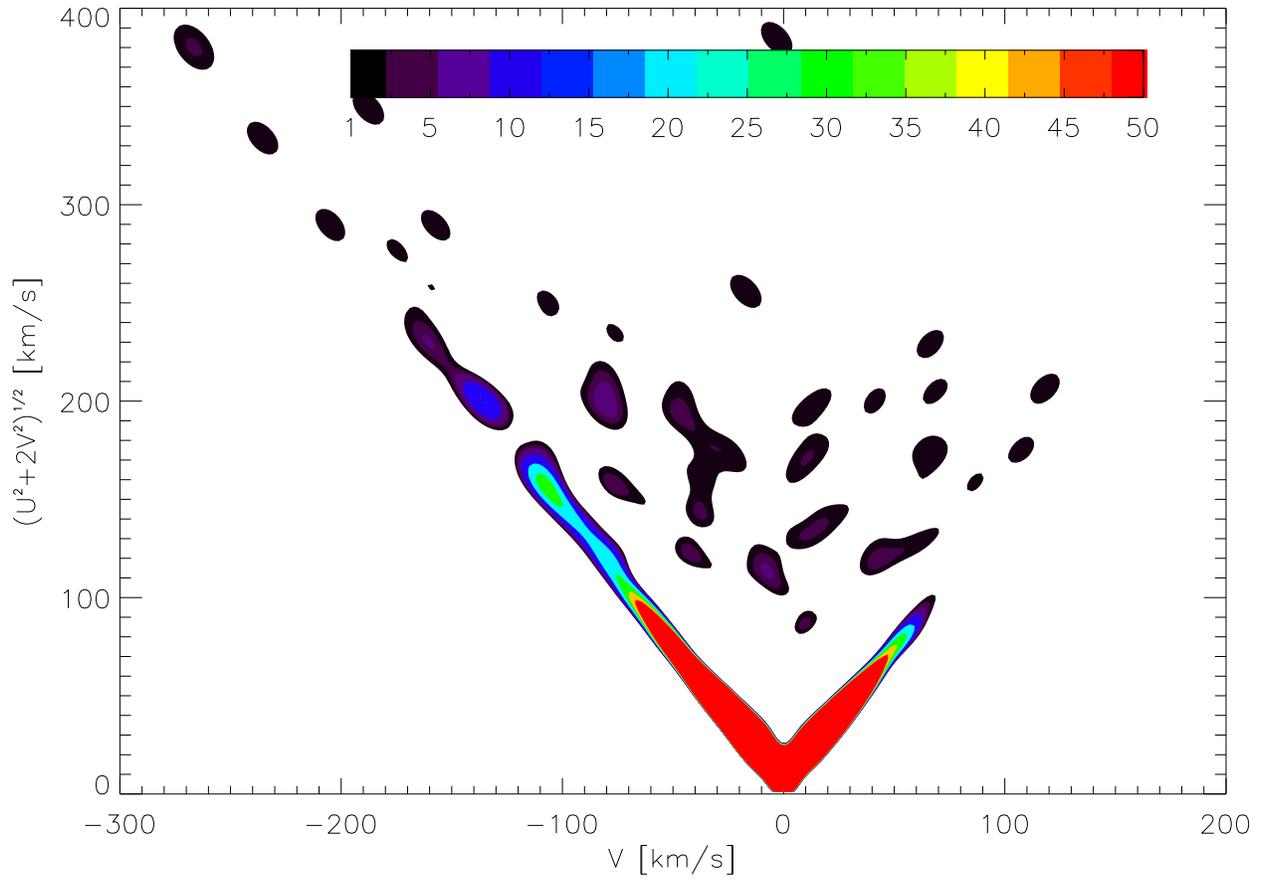}
\caption{Same as Figure~\ref{ker}, but now for one of our 250 Monte Carlo samples of 7015 stars drawn from a smooth velocity model (Section~\ref{RaD}). Due to Poisson noise, some overdensities emerge.}\label{singleWL}
\end{figure} 
\begin{figure}[!ht]
\epsscale{1.0}
\plotone{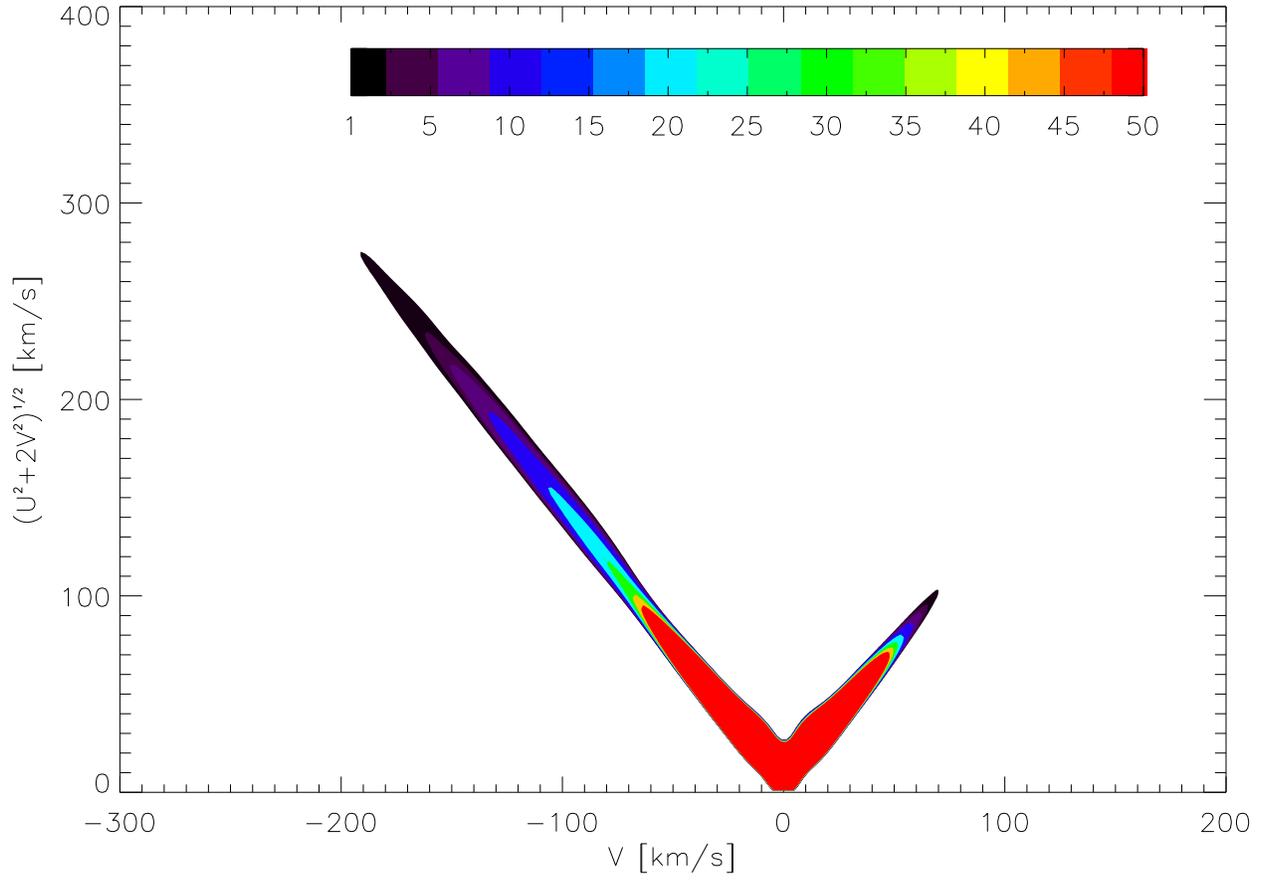}
\caption{Same as Figure~\ref{singleWL}, but now the mean value of the wavelet transform for all 250 Monte Carlo samples is shown. Only values $\geq1$ are displayed.}\label{meanWL}
\end{figure} 
\begin{figure}[!ht]
\epsscale{1.0}
\plotone{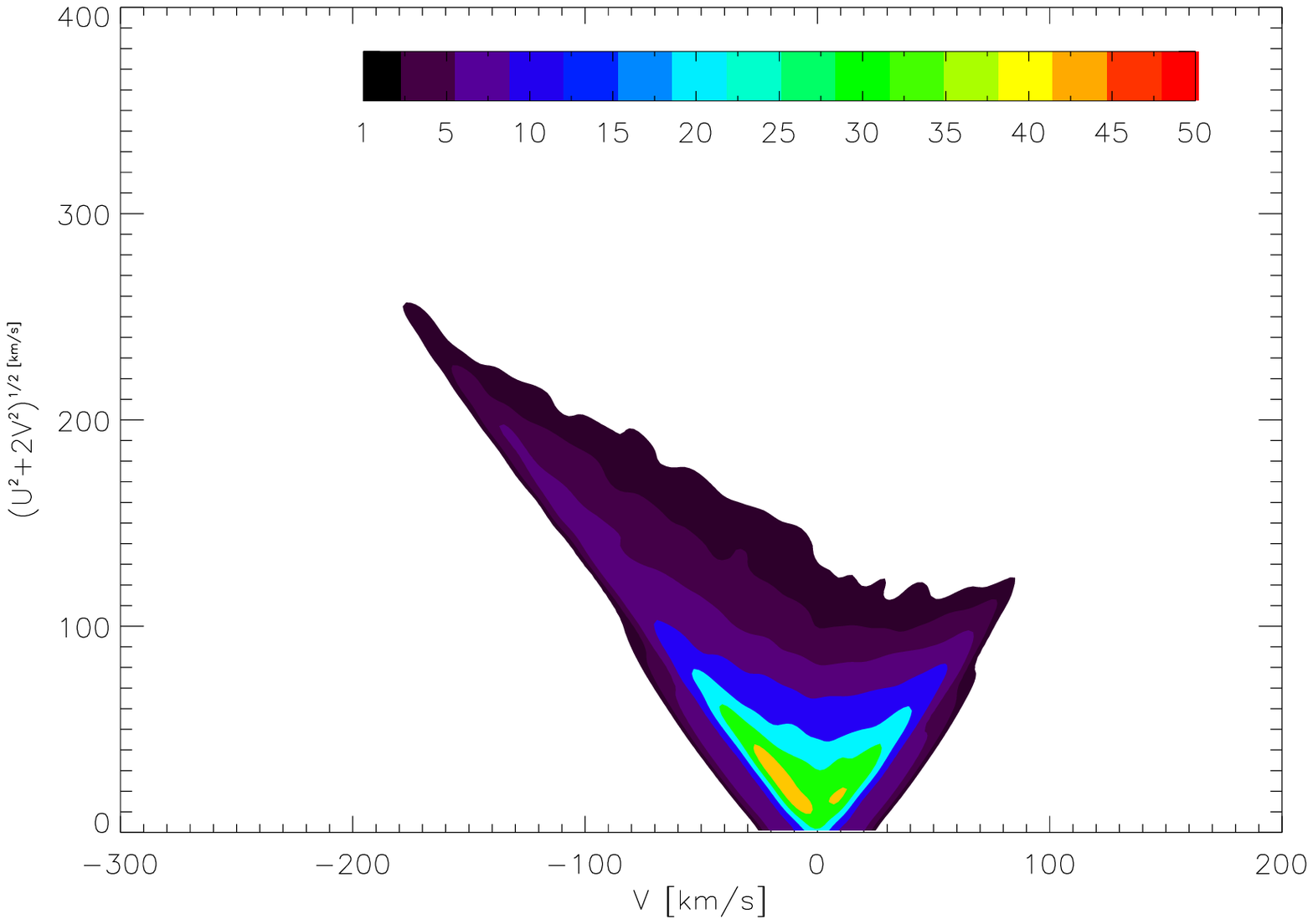}
\caption{Same coordinates as Figure~\ref{singleWL} and \ref{meanWL}, but here the standard deviation of the wavelet transform among the 250 Monte Carlo samples is shown. Outside of the contoured region, the variance is set to unity. This variance map is used to asses the significance of wavelet transform peaks in Figure~\ref{ker}.}\label{stabw}
\end{figure} 

For each MC sample of 7015 stars we binned the velocities and calculated the wavelet transform in each bin the same way as for the real sample. Figure~\ref{singleWL} displays one randomly chosen MC realization. One can clearly detect several peaks in the wavelet transform, seeming "overdensities", which must be, however, by construction due to Poisson noise. This shows that also in our real sample the possibility for such "fake" streams exists. 

We proceeded with calculating the mean value $\overline{w}_{i,j}$ and the standard deviation $\sigma_{i,j}$ of the 250 MC wavelet transforms in each bin $(i,j)$. Because many bins are not populated with stars, the standard deviation in those formally has values $\sigma_{i,j}<1$; when this was the case, we set $\sigma_{i,j}=1$. The contours of the mean and the standard deviation are shown in Figure~\ref{meanWL} and Figure~\ref{stabw}, respectively. 

While a single Monte Carlo sample shows various kinds of clumps and overdensities due to statistical fluctuations (Figure~\ref{singleWL}, and see also \citetalias{ari06}), the mean value of all wavelet transforms represents a very smooth distribution, because the clumps disappear when averaging over all samples. 

\begin{figure*}[!tb]
\epsscale{1.0}
\plotone{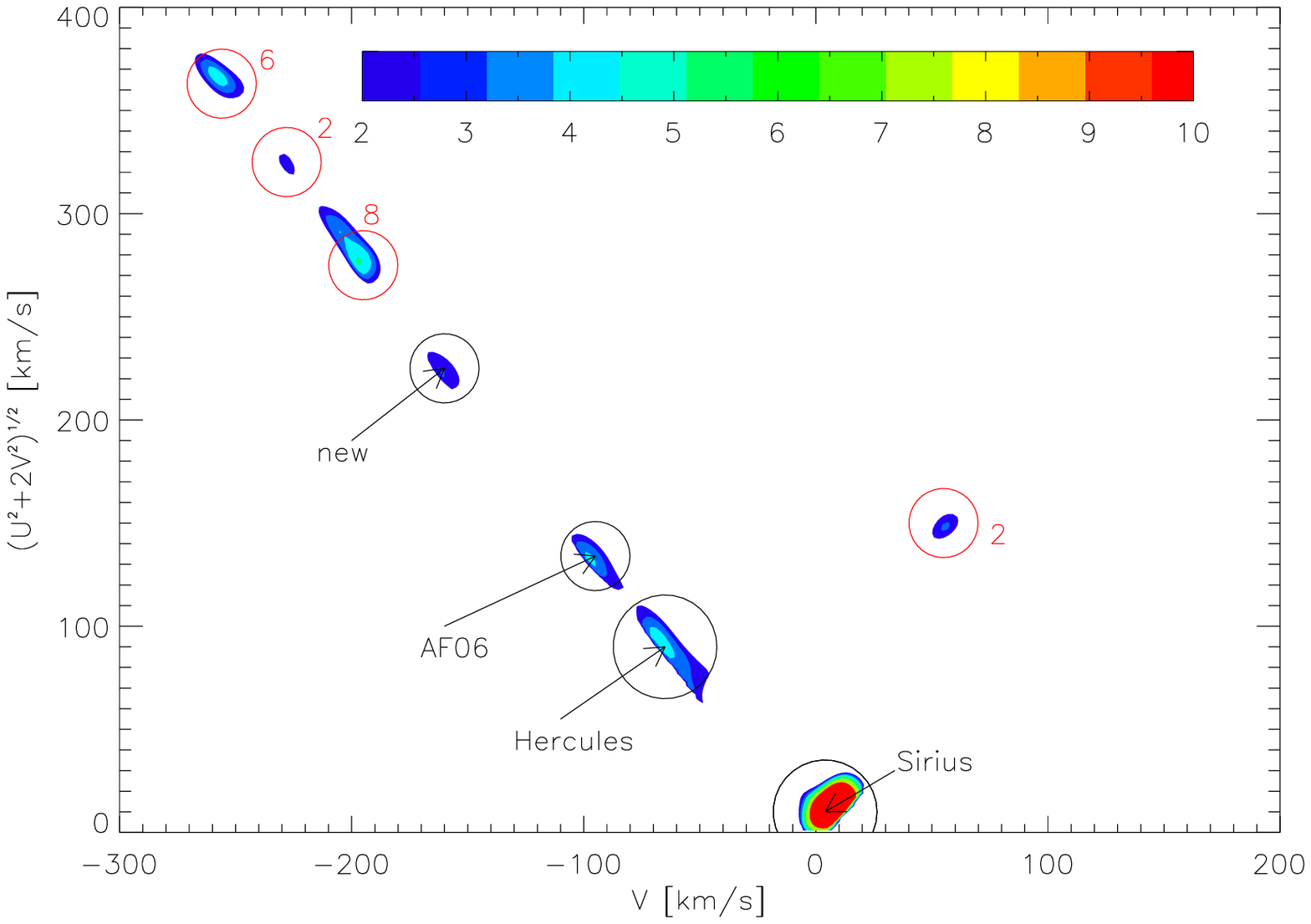}
\caption{Significance of the overdensities seen in Figure~\ref{ker}, obtained as described in the text. Note that only areas with $\sigma\geq2$ are displayed. We encircled black and labeled all features which we considere to be stellar streams. We are not making further suggestions about the four features encircled with red, but instead give the number of stars which make out these features.}\label{signifikanz}
\end{figure*}

The standard deviations $\sigma_{i,j}$ define the significance of structure in our real sample in the following way: Because we are only interested in overdense regions, i.e. pixels where the wavelet transform takes on positive values, we set the value of the wavelet transform of the data, ${w}_{i,j}^{obs}$, as well as the mean value of the wavelet transform of the 500 MC samples, $\bar{w}_{i,j}^{MC}$, to zero, whenever it is $\leq 0$. By doing so, we make sure, that the residuum $({w}_{i,j}^{obs}-\bar{w}_{i,j}^{MC})$ is $\geq 0$ in each pixel, where our "smooth" model contains no stars. By dividing the residual $({w}_{i,j}^{obs}-\bar{w}_{i,j}^{MC})$ in each pixel through $\sigma_{i,j}$ we define the significance of peaks of ${w}_{i,j}^{obs}$. For a positive stream detection, we require a significance of at least 2. Figure~\ref{signifikanz} shows the result.

\subsection{Significance of the Streams}
The first thing to mention is that in all bins where the standard deviation has a value of 1, the significance has the same value as the residual $({w}_{i,j}^{obs}-\bar{w}_{i,j}^{MC})$. In these bins, one or a few stars will result in a significant overdensity with $\sigma_{i,j} \gtrsim 2$, since for every star in the middle of a bin, the value of $w^{obs}$ increases by 2. This effect could then produce "fake" stellar streams, just because of Poisson noise in our RAVE sample and the smoothness of the combined Monte Carlo samples. Since a solution to this problem is difficult, we avoid speculating about the overdensities in bins with $\sigma_{i,j}=1$, which we encircled red in Figure~\ref{signifikanz}. We only give the number of stars that are contained in this features, as derived from the scatterplot, Figure~\ref{scatterplot}. For example the extended clump at $V\approx -200$ km s$^{-1}$ corresponds to a group of 8 stars, while the one at $V\approx -230$ km s$^{-1}$ contains only 2 stars. Further investigations with more data could provide more hints for the nature of this features.

We try to match the remaining statistically significant peaks with streams that have been already described in the literature.

Centered at $V\approx +4$ km s$^{-1}$, $|U|\approx$ 10 km s$^{-1}$, there is a big clump, significant at a level of $\gtrsim 10$. This is probably the Ursa Major/Sirius moving group \citep{dehn98, fam04}, its signal possibly amplified through the inability of our simple three-component MC models to fully match the observed thin disk distribution (see also Figure~\ref{vel-hist}). The core of this group is in the direction of Ursa Major. It's members are distributed all around the sky and can be very close (Sirius at 2.65 pc distance is a member of the group), putting the sun inside of the group \citep[ and references therein]{ban07}. The current understanding is that the Sirius stream consists not only of a cluster of coeval stars, but also of different kinds of field stars, which altogether could have been forced on similar orbits by a spiral wave gravitational field \citep{sel02,fam04,qui05}.

The elongated clump streching from $V\approx -50$ to $V\approx -75$ km s$^{-1}$ with a peak significance of $\sigma=4.5$ can probably best be described as the Hercules stream. This stream can be explained as the scattering of stars off the Galactic bar induced by the Outer Lindblad Resonance \citep{deh00}. \citet{fam04} also find a group of most likely thick disk and halo stars located at $V=-53.3\pm41.36$ km s$^{-1}$. Further, at $V=-60$ km s$^{-1}$ there exists the moving group HR1614 which is thought to be a dispersed open cluster due to its chemical homogenity \citep{egg96,silv07}. It is possible that these groups are present in our data, too, amplifying and elongating the signal of the Hercules stream. Better velocity estimates would be needed in order to clearly distinguish between these features.

The feature at $V\approx -100$ km s$^{-1}$ stands out at the $4.3\sigma$-level. It corresponds to the stream discovered independently from \citetalias{ari06} and \citet{helm06}. Most probably, this stream has its origin outside of the Galaxy and is a relic from the build-up of the Milky Way.  

\begin{figure*}[!ht]
\epsscale{1.0}
\plotone{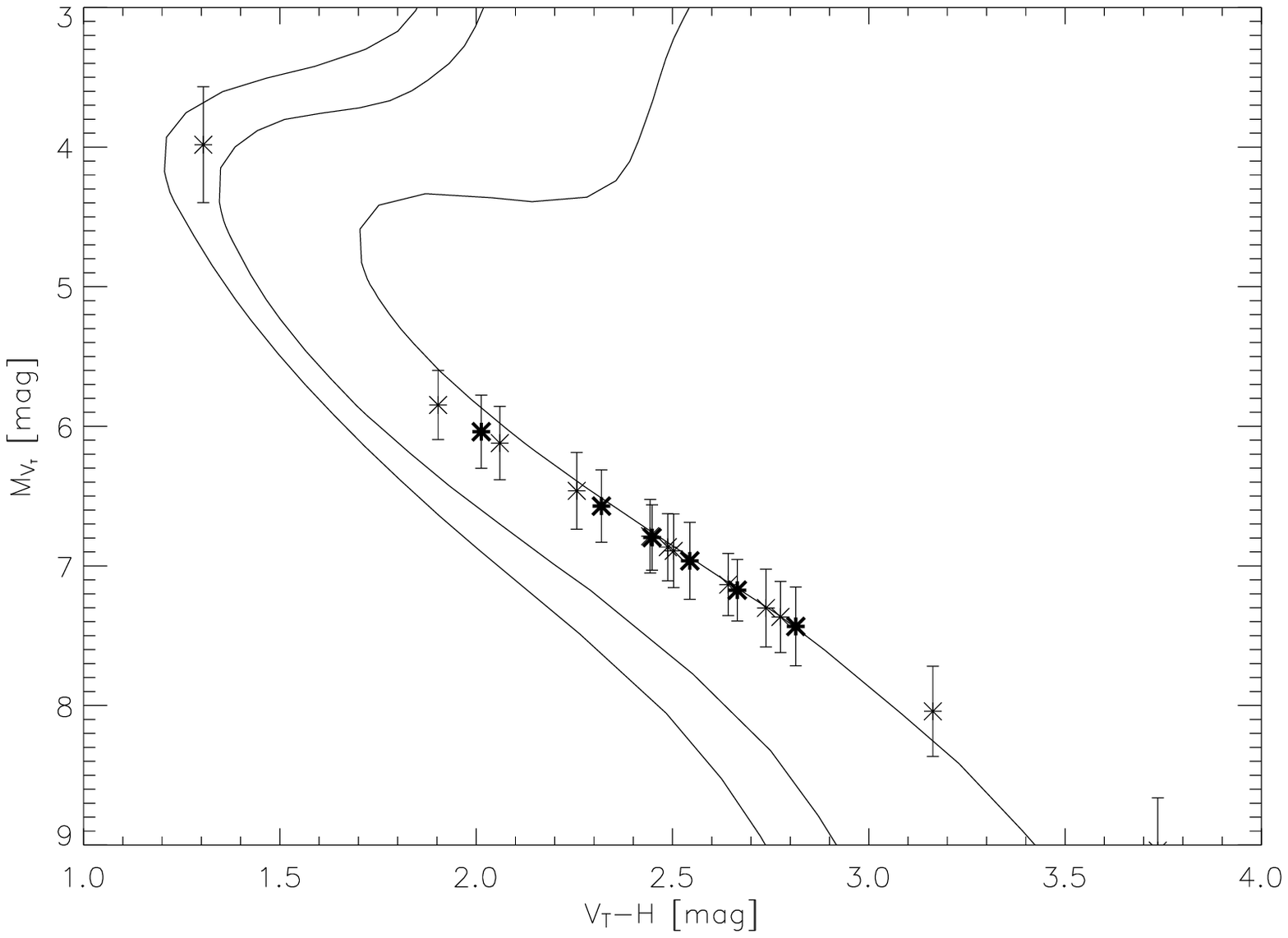}
\caption{CMD of the stars that make up the feature between $V\approx-140$ and $V\approx-180$, highlighting those in the significantly overdense region $-167$ km s$^{-1}<V<-153$ km s$^{-1}$ (thick asterisks). Also, isochrones for a 13 Gyr old population with metallicity [Fe/H]=$-1.5$, $-1.0$, $-0.5$ and 0.0 are plotted. The linearity of our adopted photometric parallax relation in the range $V_T-H > 1.9$ is clearly visible.}\label{CMD}
\end{figure*}

At a velocity of $V\approx -160$ km s$^{-1}$ there is an overdensity which stands out at the $3.0\sigma$-level in the center. A comparison with Figure~\ref{ker} reveals that this feature is probably more elongated, like one would also expect for a moving group of stars at such velocities. However, with the current sample size, our method seems not to be fully able to recover the whole range from $V = -180$ km s$^{-1}$ to $V = -140$ km s$^{-1}$ as a statistically significant overdensity. Figure~\ref{CMD} shows the stars that make up the new feature in a color-magnitude diagram consisting of $V_T-H$ color and absolute magnitude $M_{V_T}$. Stars that lie in the area with a significance greater than 2 are plotted with thick asterisks, while those that lie in the range $V = -180$ km s$^{-1}$ to $V = -140$ km s$^{-1}$ and presumably also belong to the stream have thin asterisks. We also show isochrones for a 13 Gyr old population of stars with metallicities of [Fe/H]=-1.5, -1.0, -0.5 and 0.0 (from left to right). Most of the stars seem to be consistent with having Solar metallicity, however this is a result which would be expected from the way we calculated their distances and velocities. Our assumption was that all stars are MS stars of luminosity class V, and therefore we a priori excluded the possibility to have very metal-poor (sub)dwarfs with fainter luminosities (see also footnote 2 in Section~\ref{data}). Right now, we can only wait for future data releases providing metallicity estimates to solve this issue.

In appendix~\ref{app} we investigate the Geneva-Copenhagen survey of the Solar neighbourhood \citep{nord04} with our method; although we do not determine the significance of overdensities with the help of MC sampling, we can see the same strong features that \citet{helm06} detected in their paper. It is striking that the same stellar streams can be detected in different datasets using different methods.     

Therefore - through the detection of 3 already known stellar streams in our small RAVE dataset - we have enough confidence to claim that the new feature centered on $V=-160$ km s$^{-1}$ is a strong candidate for a new stellar stream. The $W$-velocities of stars in this feature show a wide range of values, which makes it likely, that this stream was accreted long ago as tidal debris during the formation of the Galaxy. To draw more conclusions, one would need metallicity measurements, which will be available from RAVE in the future.

It is possible to give constraints on the density contrast in the streams in the Milky Way stellar halo. Therefore we estimate how many stars are needed to give a significant stream detection and divide this number by the total number of halo stars. Because from kinematics alone it is very difficult to delimit halo from thick disk stars (and even with metallicities there is no clear boundary, see \citet{chi00}), we simply take all stars outside of $\bigl(V\pm2\sigma_V\bigr)_\text{thin disk}=-5\pm42$ as halo stars and accept a contamination from the thick disk. This gives 554 stars, of which 230 (42\%) are part of the Hercules stream, 47 (9\%) belong to the AF06 stream and 7 (1\%) make up the new detected stream. We conclude that a few percent density contrast are enough to give a significant stream detection.

\section{Searching for Stellar Streams in $(U,V,W)$ and $(L_z,L_\perp)$ space}\label{App2}
The traditional concept of moving groups refers to distinct clumps in velocity space $(U,V,W)$ \citep[and references therein]{egg96}. However, as explained in Section~\ref{SSS}, this space is not immune to the effects of phase-mixing: over time, distinct clumps will disperse due to small initial differences in the oscillation periods of the constituting stars and encounters with e.g. molecular clouds. For a part of a stellar stream to be in the Solar neighbourhood at the same time means its stars must have similar $V$ motions, but the $U$ and $W$ velocities can differ by hundreds of km s$^{-1}$.

In contrast, stellar streams are able to remain as coherent features in the space of integrals of motion. Therefore, in the style of Helmi's work \citep{helm99}, we will investigate to what extent the detected streams from Figure~\ref{signifikanz} can be recovered in $(U,V,W)$ and $(L_z,L_\perp)$ space.

\subsection{Analysing the RAVE data in $(U,V,W)$ space}
\begin{figure*}[!ht]
\caption{Distribution of RAVE stars in velocity space. The upper panels show scatter plots of U vs. V and U vs. W, respectively. A lot of sub-structure can be seen. In the lower panels we show contours of a wavelet transform of the data. The convolution has been made using a Mexican hat kernel with scale parameter $a=5$ km s$^{-1}$.}\label{f13}
\end{figure*} 

\begin{figure*}[!ht]
\epsscale{1.0}
\plotone{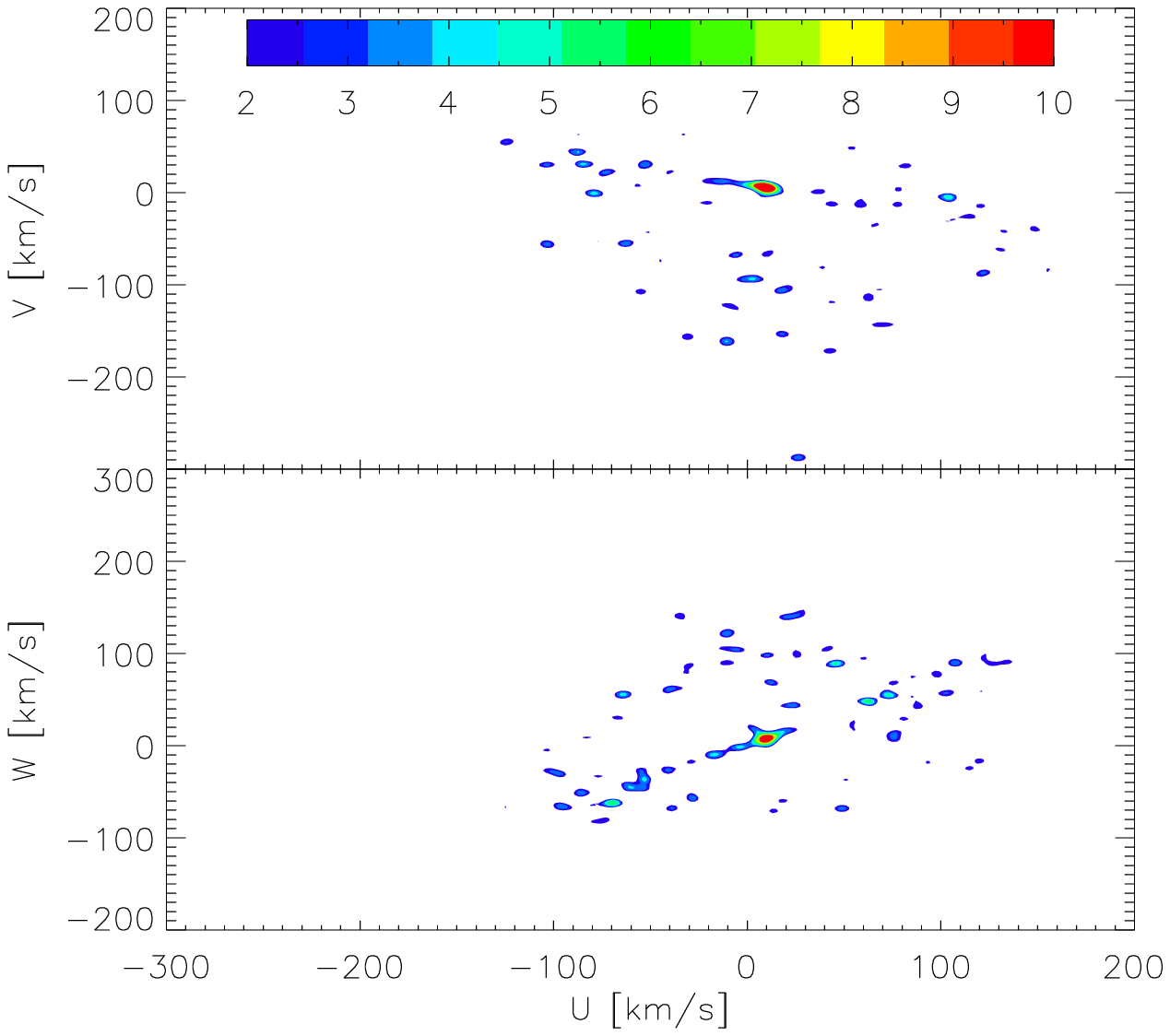}
\caption{Significance of the overdensities seen in Figure~\ref{f13}. Only areas with $\sigma\geq 2$ are displayed. Note the richness of features compared to Figure~\ref{signifikanz}, with many having a very small extension.}\label{f14}
\end{figure*}

In Figure~\ref{f13} we show the distribution of our 7015 RAVE stars in velocity space. The upper panels show how the stars are distributed with respect to their $(U,V,W)$ velocities. The distribution is far from smooth, as can be seen more clearly in the lower panels, where we display the wavelet transform of the velocity components. For the convolution we chose a Mexican hat kernel with a scale parameter of $a=5$ km s$^{-1}$ \citep[similar to] []{sku99}. The contour levels range from 1 (black) to 50 (red). The most striking features are located aound $V=0$, $U\leq=0$ and most probably can be attributed to dynamical thin disk streams like the Sirius moving group. The elongated shape along the $U$-axis is typical for stellar steams and is caused by the fact that we observe stars with slightly different orbital phases. The classical moving groups Sirius at $(U,V)\approx(+6,+4)$ km s$^{-1}$, Hyades-Pleiades at $(U,V)\approx(-25,-15)$ km s$^{-1}$ and Hercules at $(U,V)\approx(-20,-50)$ km s$^{-1}$ can be found with good agreement with the values given by \citet{fam04}. The incomplete sky coverage of our sample as well as the already described effects of phase mixing make it hard to speculate about the meaning of the many other clumps. We nevertheless performed a statistical Monte Carlo analysis analog to that described in Section~\ref{Analysis} using 200 Monte Carlo simulations. Figure~\ref{f14} provides the significance map of the velocity space, showing all overdensities from Figure~\ref{f13} that have a significance of at least 2. 

Compared to $(V,\sqrt{U^2+2V^2})$ space (c.f. Figure~\ref{signifikanz}) there are more features present, but most of them are very small and at lower significance levels. Besides the prominent red feature around $(U,V)\approx(0,0)$, which can be attributed to the Sirius stream, it is not easy to associate the identified streams from Figure~\ref{signifikanz} with clumps in the $(U,V)$ or $(U,W)$ diagrams, respectively. Because old moving groups are not expected to remain coherent as small clumps in velocity space, we likely over-resolve substructure due to the small deviations of our Monte Carlo model from the data. These deviations can also lead to a wash-out of the banana-shaped structures in $(U,V)$ that are expected for old tidal streams \citep{helm06}. Large errors in the velocities can wash out these structures, too, but in our case the errors are small enough and should allow their detection. The fact that a clear classification of particularly old stellar streams from the $(U,V)$ and $(U,W)$ distributions is not possible, and that many small features are even less significant than the overdensities in $(V,\sqrt{U^2+2V^2})$ space lead us to conclude that $(U,V,W)$ space is not suited well to detect moving groups.  
 
\subsection{Analysing the RAVE data in $(L_z,L_\perp)$ space}
In a spherical static potential the two components of the angular momentum $L_z$ and $L_\perp=(L_x^2+L_y^2)^\frac{1}{2}$ are integrals of motion. In this case, tidal and dynamical streams remain as coherent clumps in $(L_z,L_\perp)$ space. The asumption of a spherical potential seems valid for large heights above or below the disk which are reached only by halo stars, but not necessarily in our case of nearby stars with planar orbits. These stars move under the influence of a flattened potential and the radial action integral $J_R$ would be better suited as a conserved quantity. The space of integrals of motion has been shown to be useful for the detection of stellar halo streams in recent studies \citep{helm99,helm00,chi00}. 

\begin{figure*}[!ht]
\caption{Distribution of RAVE stars in $(L_z,L_\perp)$ space. The upper panel is a scatter plot of the data. In the lower panel we show contours of a wavelet transform of the data. The convolution has been made using a Mexican hat kernel with scale parameter $a=45$ km s$^{-1}$.}\label{f15}
\end{figure*} 

In Figure~\ref{f15} we show the distribution of our RAVE stars in $(L_z,L_\perp)$ space, again as a scatter plot (upper panel) and after wavelet transformation using a Mexican hat kernel with scale parameter $a=45$ kpc km s$^{-1}$. For stars near the sun $L_z\simeq R_\odot (V+V_{LSR})$, so we expect to identify again the overdensities from Figure~\ref{ker}. 

A comparison shows first of all that $(L_z,L_\perp)$ seems to be much more structured than $(V,\sqrt{U^2+2V^2})$ space, but this is partly due to the applied chosen scale parameter $a$ in the kernel function. The other reason is that now all three velocity components go into the calculation of the angular momentum components. Also, there are no "forbidden regions" as in the case of $(V,\sqrt{U^2+2V^2})$ space. 

Some overdense regions in the lower panel can be related to overdensities in $(V,\sqrt{U^2+2V^2})$ space. For example, the "bump" in the distribution in Figure~\ref{ker} at $V\approx -15$ km s$^{-1}$ is also present in Figure~\ref{f15} at $L_z\approx 1640$ kpc km s$^{-1}$. Also, the Hercules stream is present at $L_z\approx 1350$ kpc km s$^{-1}$. 

\begin{figure*}[!ht]
\epsscale{1.0}
\plotone{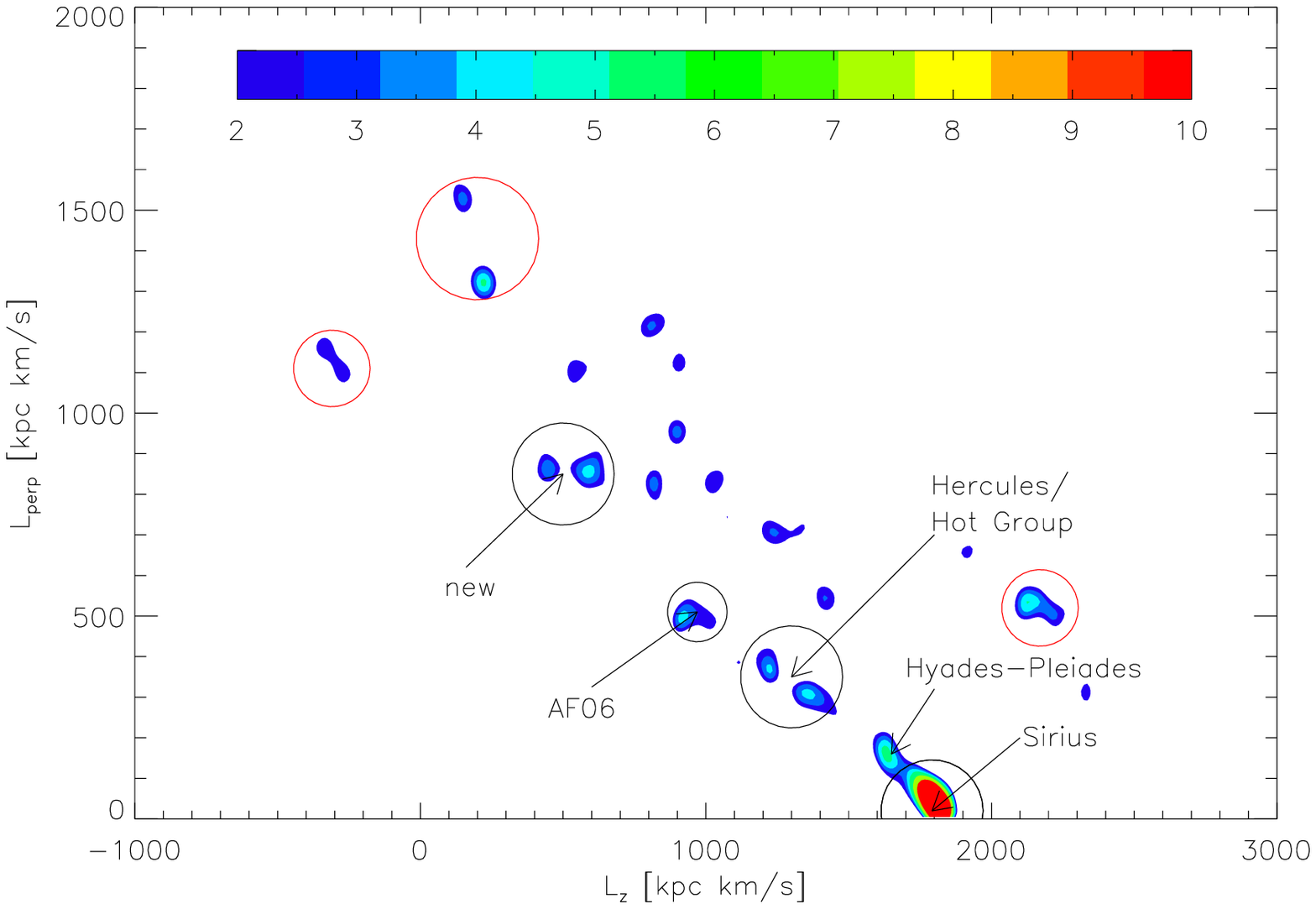}
\caption{Significance of the overdensities seen in Figure~\ref{f15}. Only areas with $\sigma \geq2$ are displayed. Black circles mark features which we connect to the detections of Figure~\ref{signifikanz}; there is further evidence for the now significant Hyades-Pleiades group. Red circles mark features which can be connected to doubtful features in Figure~\ref{signifikanz}. Note that for stars near the sun $L_z\approx R_\odot V$, were we take $R_\odot=8$ kpc. } \label{f16}
\end{figure*}

We proceeded to investigate which overdensities proof significant in the same way as before by using 200 MC samples that were drawn from the 3-component Galaxy model described in Section~\ref{Analysis}. After building the mean value and standard deviation of their wavelet transforms we derived the significance map, which is shown in Figure~\ref{f16}.

As in the case of $(U,V,W)$ space the number of features with significance $\sigma\geq 2$ is larger than in Figure~\ref{signifikanz}. We think this is caused by the extra-information of $W$ velocity required for the MC samples. Some of these features are very small and again belong to regions of phase space that are sparsely sampled, but others seem to be comparable to the already known ones. We encircled all features which correspond to the detections of Figure~\ref{signifikanz} with black circles for the appointed streams or with red circles for the doubtful ones. Additionally, there is a new feature present at $L_z\approx 1650$ kpc km s$^{-1}$ which we identify as the Hyades-Pleiades group. A hint on the presence of this group has already been seen as a "bulge" in the distribution of stars in $(V,\sqrt{U^2+2V^2})$ space (Figure~\ref{ker}). It is interesting that some distinct features in $(L_z,L_\perp)$ space (e.g. the new stream) project onto the same feature in $(V,\sqrt{U^2+2V^2})$ space. Furthermore, when we select stars from the fastest rotating clumps (Hyades-Pleiades, Sirius and the red encircled clump on the right side in Figure~\ref{f16}) we find that they span a wide range of $\sqrt{U^2+2V^2}$ values. On the other hand we also find that clumps in $(V,\sqrt{U^2+2V^2})$ space are not necessarily projected onto clumps in $(L_z,L_\perp)$ space, indicating that $L_\perp$ is not conserved for their motion. Therefore, we conclude that $L_\perp$ compared to $\sqrt{U^2+2V^2}$ is not a good choice as an approximate integral of motion if the disk's potential dominates the movement of the stars.

We note as a main result that the significance of the detected features is comparable, no matter which space we use for the stream search. As explained before, we expect the significance of the detections to go up with a larger sample size. For best results, it is probably good to combine the searches in $(L_z, L_\perp)$ and $(V,\sqrt{U^2+2V^2})$ space to find stellar streams on both inclined and planar orbits. Also, the significance is influenced by the statistical method which is used to identify overdense regions. Our MC method has the drawback that a good Galactic velocity distribution model is needed in order to avoid "fake" detections in regions where the data points are rare. Maybe a combination of different analyses can also enhance the detection efficiency.

\section{Conclusions}
To explore RAVE's practical potential for finding kinematic substructures (streams, or moving groups) in the Solar neighbourhood, we have studied a sample of 7015 nearby stars from the RAVE first public data release, which was selected to have acceptable distance and velocity estimates. Distances have been derived from a photometric parallax relation, based on Hipparcos MS stars using the $V_T$ and $H$ bands. For our sample, we required a distance error better than 25\%, while the mean velocity errors in $U$ and $V$ are $\approx -5$ km s$^{-1}$, respectively. 

To search for streams, we plotted $V$ velocities against the quantity $\sqrt{U^2+2V^2}$, which essentially corresponds to angular momentum $L_z$ against orbital eccentricity $e$. Streams should form clumps in this projection of phase space. To identify such clumps more obvious, we applied a wavelet transform with a modified two-dimensional Mexican kernel as the analysing wavelet. Several overdensities are visible in this sample, presumably corresponding to stellar streams. We tested whether these clumps are real or due to Poisson noise using 250 MC simulations drawn from a three component Schwarzschild distribution. While it is very difficult to draw conclusions about the amount and significance of sub-structure present in the very few halo stars, we detected three already known stellar streams and one candidate for a newly discovered stream. The latter is present as a broad feature in the range $-180\leq V\leq -140$, centered at $V\approx-160$ km s$^{-1}$ and from its kinematics would belong to the stellar halo population. Its other velocity components (high $W$ velocities with a mean of $\overline{W}=117\pm26$ km s$^{-1}$, $\overline{U}=6.7\pm55$ km s$^{-1}$) make it likely that this stream is part of the tidal debris from an accreted satellite rather than a dynamical resonance. The kinematics, however, could be somewhat biased, because we did not correct for metallicity effects in the photometric parallax relation. The other moving groups are the Sirius stream at $V\approx+4$ km s$^{-1}$, the Hercules stream centered at $V\approx-65$ km s$^{-1}$\citep{dehn98,fam04}, and the stream discovered by \citetalias{ari06} and \citet{helm06} at $V\approx-100$ km s$^{-1}$.

We are missing a detection of the Hyades-Pleiades and Arcturus streams which were still present as peaks in the wavelet transform of the data in Figure~\ref{ker}. Here, the difference $({w}_{i,j}^{obs}-\bar{w}_{i,j}^{MC})$ is too small to make the peak in ${w}_{i,j}^{obs}$ statistically significant. 

However, we were able to detect the Hyades-Pleiades moving group in velocity and angular momentum space, which we used to complement our stream search. By comparing the detections in $(U,V,W)$, $(L_z,L_\perp)$ and $(V,\sqrt{U^2+2V^2})$ space we showed that i) the significances of the features were comparable, ii) that velocity space is sub-optimal for detecting stellar streams, but may be useful as an additional information on the age and origin of a stream (tidal streams look differently from dynamical streams, old streams disperse) and iii) that $L_\perp$ is a poor approximate integral of motion compared to $\sqrt{U^2+2V^2}$ if the stars move under the main influence of the disk's potential.

The fact that only a fraction of the DR1 stars were enough to find significant substructure in the Solar neighbourhood shows the power of the method we used. Subsequent data releases will not only enlarge the sample of stars and eventually allow the detection of further clumps in phase space, but also provide measurements of $\log g$, $T_{eff}$ and $[M/H]$. The full dataset will include these measures for up to one million stars \citep{stei06}. Simply speaking, and neglecting an increase in the spatial coverage, an enlargement of the sample size by a factor of N leads to N times more stars in a stellar stream, while the standard deviation of the MC samples with be increased by a factor of $\sqrt{N}$ (Poisson noise). The mean value of the wavelet transform of the MC samples stays the same, so that the significance of the features will be $N/\sqrt{N}=\sqrt{N}$ times as high. In this way we can hope to detect more features and put further constraints on the exact nature and origin of the detected streams.  

\appendix
\section{Streams in the Geneva-Copenhagen survey}\label{app}
We show the performance of our method on the data taken from the Geneva-Copenhagen survey of the Solar neighbourhood \citep{nord04}. \citet{helm06} searched for stellar steams in velocity and APL space in these data. While the dynamical streams (Sirius, Hyades-Pleiades and Hercules) can be seen as clumps in velocity space, a detailed statistical analysis of the APL space revealed around 10 further overdensities oriented along two to three segments of constant excentricity. \citet{helm06} divided the stars in the overdense regions into three groups based on their metallicities. There seems to be a relation between these three groups, the Arcturus stream and the stream detected by \citet{ari06}. 

\begin{figure*}[!ht]
\epsscale{1.0}
\plotone{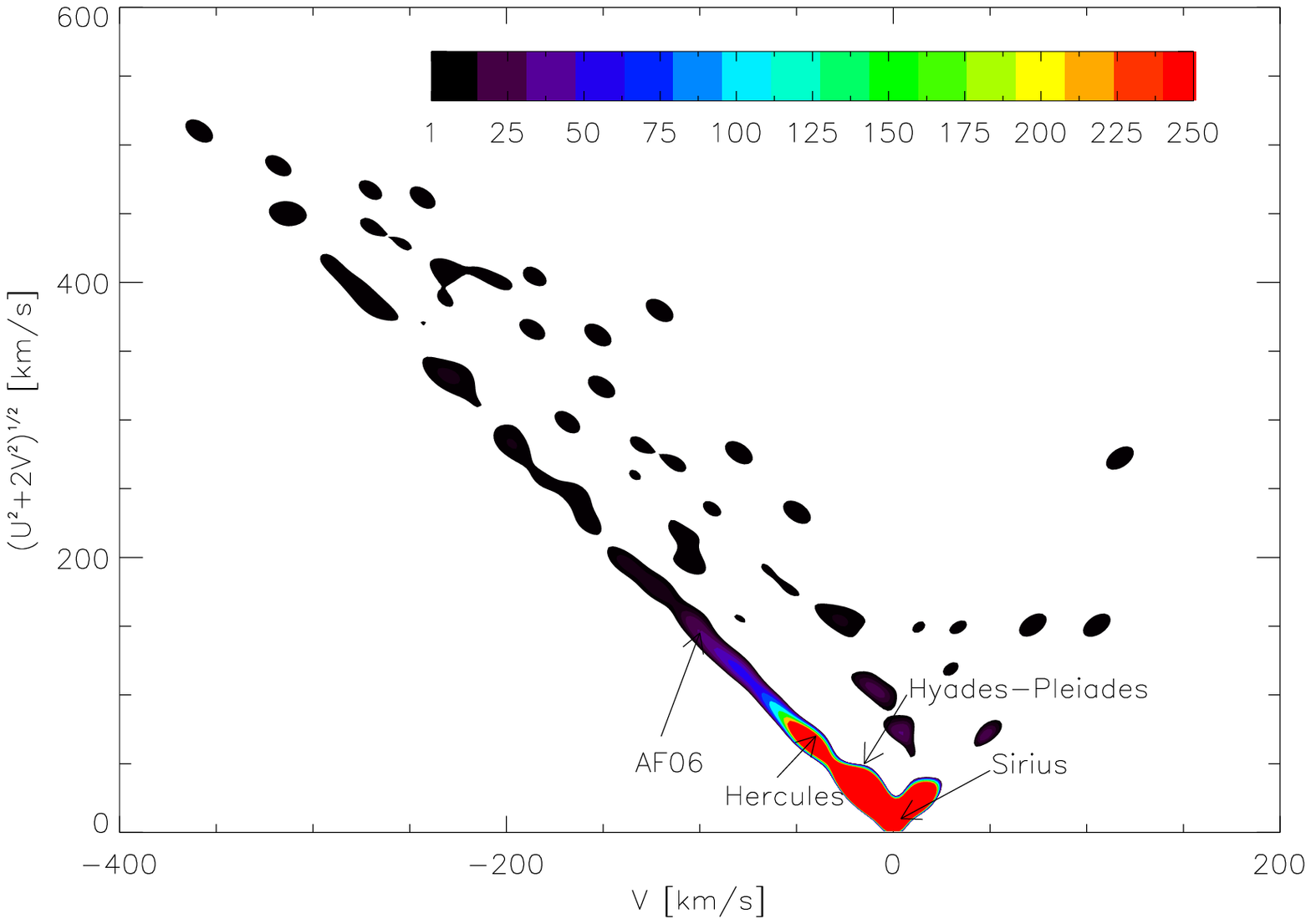}
\caption{Contours of the wavelet transform of 13440 stars from the Geneva-Copenhagen survey plotted in $V$ vs. $\sqrt{U^2+2V^2}$ (most of which are not statistically significant). The features that have been described by \citet{helm06} have been labeled.}\label{Nord}
\end{figure*} 

In Figure~\ref{Nord} we just show the distribution of the 13240 Geneva-Copenhagen stars in $V$ vs. $\sqrt{U^2+2V^2}$, after convolution with the analysing wavelet given by Equation~\eqref{anawavelet}.\footnote{\citet{ari06} also did this analysis, but using the presumably not optimal Mexican hat analysing wavelet.} The most striking features are those that have also been found by \citet{helm06}. Additionally, a comparison with our RAVE sample (Figure~\ref{ker}) seems to suggest that there is much more sub-structure present in the Geneva-Copenhagen survey, although it contains only $\sim2$ times more stars. We find that 10582 of the 13240 stars are included in these features with a value of the wavelet transform $w_{i,j}\geq1$. That makes up for 79.9\% of the sample. Our RAVE sample contains 7015 stars, of which 5601, or 79.8\%, constitute to regions with $w_{i,j}\geq1$. So it seems that the size of the sample is the main reason that in the Geneva-Copenhagen survey we find more distinct overdensities. Furthermore, if we relate the number of patches above the V-shaped $U=0$ line (30 in the Geneva-Copenhagen, 14 in the RAVE sample) to the number of stars in the sample, we find a similar relation; the larger number of stars increases the probability to find outliers in less populated regions. Finally, stars in the Geneva-Copenhagen survey have accurate trigonometric parallaxes from Hipparcos with relative errors $(\sigma_\pi/\pi)$ better than 10 per cent, in contrast to the less acurate photometric parallaxes of our RAVE stars. This implies less accurate velocities that tend to smear out small nearby features in phase space.

Although it's unlikely that the majority of the small features in Figure~\ref{Nord} has any statistical significance, this example still shows that a large sample size together with good distance estimates is crucial for detecting stellar streams in phase space. Still, our photometric parallaxes seem to be good enough and our sample large enough for significant stream detections. Without investigating the statistical significance of the features in Figure~\ref{Nord}, we leave this example as further evidence that stellar streams will be detectable as clumps in $(V,\sqrt{U^2+2V^2})$ space.

\clearpage
\end{document}